\documentclass[preprint,aps,prd,amsmath,showpacs,nofootinbib]{revtex4}
\usepackage{graphicx}

\newcommand{\beq}{\begin{equation}}
\newcommand{\eeq}{\end{equation}}
\newcommand{\bea}{\begin{eqnarray}}
\newcommand{\eea}{\end{eqnarray}}

\newcommand{\Ref}[1]{(\ref{#1})}

\begin{document}

\title{Einstein's hole dilemma and "gauge" freedom.}

\author{Sergey M. Kozyrev}
\email{Sergey@tnpko.ru} \affiliation{Scientific center for gravity
wave studies ``Dulkyn'', Kazan, Russia}

\author{Rinat A. Daishev }
\email{Rinat.Daishev@ksu.ru} \affiliation{$^1$ Institute of
Physics, Kazan Federal University, Kremlevskaya str. 18, Kazan
420008, Russia}

\begin{abstract}
The issue of the physical equivalence between the different
coordinate system in Einstein theory is revised. "Gauge" fixing
influences results of measurements and physics are different in
two different coordinate system. Spacetime metric generated by
static spherically symmetric distribution of matter can be matched
with wide family of vacuum solution and the exterior spacetime
geometry could not be deduced directly from the interior perfect
fluid solution, without reference to a "gauge" fixing or
viceversa. The property of sollutions in general relativity is
indeed an observer dependent concept.
\end{abstract}


\maketitle

\section{Introduction}
In the usual description in the frameworck of general relativity
our spacetime have a pseudo-riemannian geometry.
However, this an arbitrary
chose and the general theory of relativity may be formulated in
the language of Weyl geometry \cite{Romero}, for example.

General relativity describe the universe as grounded on
differentiable arbitrary manifold $M^4$ enveloped by a principal
bundle formed of isometric representations of a finite continuous
Poincar\'{e} group. Einstein's principle of general relativity
asserts the invariance under general coordinate transformations of
the actions integral grounded on a $M^4$ manifold parameterized by
variables  $x^\mu, \mu = 0,1,2,3$.

According to the theory of manifolds one can always introduce the
coordinate chart through a map from an open set in the manifold
$M^4$ to an open set of $R^4$. This approach adopt an
interpretation of the variables $x^\mu$ as mere mathematical
parameters, devoid of any geometrical significance. The parameters
$x^\mu$ then do not classify an operationally well-defined
position in space and time, although they can be regarded as
defining a chart on an abstract manifold. Such a manifold should
not, however, be confused with the space of all events, which
requires the presence of physical fields for its very definition
\cite{Westman}.

There is also Einstein's famous 'hole argument' in general
relativity which asserts that the notion of a space-time point (in
a manifold) has no physical meaning in a theory that is invariant
under the group of space-time diffeomorphisms \cite{Einstein}.
Evidently, the use of coordinates is optional, and that one could
adopt a coordinate free description of same manifold $M^4$, in
this approach, the manifold points cannot correspond to
operationally well-defined events. Thus in general relativity
coordinates in manifolds are physically meaningless before
specifying the metric tensor though they designate a particular
point of the underlying mathematical manifold.
\section{Variational principle}

As is well known field equation and conservation low of the
relativity theory can be obtained from principle of least action.
The same principle is the basis of the general relativity
\begin{eqnarray}
S & = & \int\sqrt{-g}R(g^{\mu\nu},
\partial_\lambda g^{\mu\nu})d^4x + \int\sqrt{-g}\,L^M(g^{\mu\nu},
\partial_\lambda g^{\mu\nu}, {\boldsymbol\Psi}, \partial_\lambda {\boldsymbol\Psi})\,d^4x \,, \label{action}
\end{eqnarray}
where respectively we have the action integral of the geometry and
the action integral of matter; $R$ is the Ricci scalar a function
of $g^{\mu\nu}$ and their partial derivatives, $L^M$ the
Lagrangian density of the matter as a functional of the metric
tensor and a set of non-gravitational fields ${\boldsymbol\Psi}$.

Note that in abstract manifold the Ricci scalar and tensor
$g_{\mu\nu}$ lose their geometrical meaning they had in a
spacetime and now can be viewed only as a "source" for the metric.
An unsatisfactory feature of general relativity is that the
components of Lagrangian do not have any direct physical
interpretation. Moreover,  Albert Einstein in 1923 assumed a
priory that both a metric and connection must be chosen, from the
beginning, as dinamical variables. Note that an affine connection
is not uniquely defined by the Lagrangian structure and can be at
most an independent postulat of theory \cite{Brans}. In this case
the point dependent property of manifolds is linked with the fact
that the units for measure of underlying geometry will be running
units. For example, it could be a theory based on Einstein Hilbert
action but endowed with space time of Weyl integrable structure
\cite{Matos}.

The general relativity appears as a theory in which the gravity is
described simultaneously by two fields the metric tensor and the
matter fields, the latter being an essential part of the
geometrical property of spacetime  (emerged after solution of the
field equations) manifesting its presence in almost all
geometrical phenomena, such as curvature, geodesic motion and so
on. 

While the gravitational interaction are described by a doublet
constituting of a metric tensor and a matter fields, the important
aspects of general relativity is connected with conformal
symmetry. The matter terms of the Lagrangian density contain the
connection and hence a part of dynamical description of gravity,
that is invariant in form under the conformal "rescaling" of the
metric. This conformal symmetry is sufficient to guarantee the
invariance of the Lagrangian under arbitrary changes of variables.
It is then evident that the total Lagrangian may contain the terms
for one or more physical fields, so that we can shift it by
changes of variables from geometric action integral $ S^G$ to
matter action integral $ S^M$.

Note that the limit case is a flat space, in this frame, similarly
as in electrodynamic the $g^{\mu\nu}$ are 10 gravitational
potentials and directly interacts with matter. The Cristoffel
symbols are the gravitational field strength. Since the
gravitational potentials are not observable quantities it has no
direct physical interpretation in general.

  It was first recognized by Weyl \cite{WEYL1} and fully developed
by Cartan \cite{CARTAN1} - \cite{CARTAN3} that in such a systems
one can always make a field redefinition (a change of variables)
to another arithmetization of manifold via conformal mapping

\beq
 g_{\mu\nu}(x)\rightarrow
e^{2\alpha(x)}\widetilde{g}_{\mu\nu}(x).
\label{Conf} \\
 \eeq
where $\alpha(x)$ is a differentiable real function of manifold
parameters.

 The tensor calculus on these manifolds is enriched by
new properties, which are completely explained by the Weyl
transformations of a few basic quantities. Taking as
fundamental--tensor variation the finite transformation
(\ref{Conf}), we obtain by the Weyl transformations
\begin{eqnarray}
\label{Gammavariations} \!\!\!\!\!& & \Gamma^\lambda_{\mu\nu}
\rightarrow \tilde \Gamma^\lambda_{\mu\nu}=
\Gamma^\lambda_{\mu\nu} + \delta^\lambda_\nu
\partial_\mu \alpha + \delta^\lambda_\mu \partial_\nu \alpha -
g_{\mu\nu}\partial^\lambda \alpha
\,; \\
\label{RmunutotildeRmu}
 \!\!\!\!\!& & R_{\mu\nu}\rightarrow \tilde R_{\mu\nu}= R_{\mu\nu}- 2
\bigl[ g_{\mu\nu} g^{\rho\sigma} \partial_\rho \alpha
\partial_\sigma \alpha
-\partial_\mu \alpha \partial_\nu \alpha +\nonumber \\
& &\qquad\qquad\qquad\,\,  \nabla_\mu\partial_\nu
\alpha\bigr]-g_{\mu\nu} \nabla_\lambda \nabla^\lambda \alpha\,;\\
\label{RtotildeR} \!\!\!\!\!& & R \rightarrow \tilde R =
e^{-2\alpha}\bigl[R-6 g^{\rho\sigma} \partial_\rho \alpha
\partial_\sigma \alpha-6 \nabla_\lambda \nabla^\lambda \alpha \bigr]\,;
\end{eqnarray}
where $\delta_\mu^\nu$ is the Kronecker delta function and
$\nabla_\mu$ is respectively the covariant differential operator
constructed out of $g_{\mu\nu}(x)$.
Eqs.(\ref{Gammavariations})--(\ref{RtotildeR}) describe the
structural changes of the basic tensors of the differential
calculus.

This change of variables lead us to extension of Riemann
connection by Weyl transformation implies the extension of the
Poincar\'{e} group, to the conformal group. This is possible
provided that the Weyl transformation act on any field
representation $\Psi$ according to the low
\begin{equation}
\Psi(x)\rightarrow \tilde\Psi(x) =
e^{w_\Psi\alpha(x)}\Psi(x)\,,\nonumber
\end{equation}.

It is important to note that the two geometrical structures, the
metric and arithmetization, are fundamentally independent
geometrical objects. Thus the theory can be expressed in terms of
infinite number of related charts.

This approach consists of introducing an extra geometrical entitys
in a manifold a 1-form field, for example, in terms of which the
Riemannian compatibility condition between the metric $g$ and the
connection $\Gamma$ is redefined. Then a group of transformations
which evolves both $g$ and "matter field", is defined by requiring
that under these change of variables the new compatibility
condition remain invariant. In a certain sense, this new
invariance group include the conformal transformation as subgroup.
In other terms this "ordinary matter" may be represented in
disguise in infinite number of way as a gravity component or as a
conformally invariant matter fields. Once matter has been coupled
to gravity in a frame one has a freedom to make a change of
variables to any other frame. So property of Einstein Hilbert
action in general relativity is indeed an observer dependent
concept.

\section{Field equations}

We can derive the field equation from the variational equation
\begin{equation}
\frac{\delta S}{\delta g^{\mu\nu}(x)}\equiv \frac{\delta (S^G
+S^M)}{\delta g^{\mu\nu}(x)}=0\,,\nonumber
\end{equation}
stating the invariance of the total action under change of
variables. Since we have
\begin{equation}
R_{\mu\nu}- \frac{1}{2}g_{\mu\nu}R =T^M_{\mu\nu}\,,\label{EquE}
\end{equation}
this equation is usually understood as the equation which relates
spacetime geometry to the distribution and motion of matter field.
Starting from the Hilbert - Einstein equation, we must bear in
mind that inherent in this equation is the coordinates $x$
represent only a certain manifold, fixed by selectable
arithmetization. In addition there is the Bianchi identity a
simple consequence of these symmetry properties is that the field
equations alone are not enough to determine a gravitational
system, while these equations are a set of 6 nonlinear partial
differential equations for the 10 metric components.  Einsteins
equations determine the solution of
a given physical problem up to four arbitrary functions.

One of the largest concentrations of literature within the area of
relativistic gravity theories is interpretations of exact
solutions of field equations. Some of them are discovered at the
early stage of development of relativistic theories, but up to now
they are often considered as equivalent representations of some
"unique" solution.

Evidently, a structure of space-times is mathematically
represented by Einsteins equations \Ref{EquE} and four co-ordinate
conditions \cite{Temchin}, which considered independent of the
action
\begin{equation}
C(x)g_{\mu\nu} = 0,\label{FeqI}
\end{equation}
 where  $C(x)$ - some algebraic or differential operators.
Thereby for any four of components $g_{\mu\nu}$ emerge the
relations with remaining six and, probably, any others, known
functions. Certainly, equations \Ref{FeqI} cannot be covariant for
the arbitrary transformations of independent variables, and
similarly should not contradict Einstein's equations or to be
their consequence. Moreover, these four equations will not be
transformed according to any rules, but simply replaced by hand
with the new. From geometrical point of view one has to introduce
an additional mathematical structure - describing some specific
principle of construction of space-time model is responsible for
measuring the distances - the "gauge". In general relativity
"gauge" and coordinate transformation means the same thing.  On
the other words "gauge" is a rule for reception of "coordinate
system" on a single manifold $M$ (e.g. harmonic, isotropic,
curvature coordinates). This "gauge" is the unphysical degree of
freedom and we must fix the "gauge" or extract some invariant
quantities to obtain physical results \cite{Nakamura}.

The unknown components of metric tensor $g_{\mu\nu}$  are
determined from the solutions of Einstein's field equations.
Consequently, the geometrically interpreted co-ordinate system of
obtained space-time and any relationship it derives from equations
\Ref{EquE}, \Ref{FeqI} emerge a posteriori \cite{Temchin}.
Moreover, property of this co-ordinate system will depend from
initial and boundary conditions for \Ref{EquE}, \Ref{FeqI}. An
intriguing consequences of the above discussion is the "gauge"
freedom can be expected in relation with some connection to
problems in quantum physics. Generally speaking, occurrence of the
observer ("gauge" fixing) influences results of measurements and
physics are different in two different "gauges" \cite{Gullstrand}.

The distinct geometrical and physical picture of the same
phenomena may arise in a different "coordinate systems". The
physical content of this point of view can be stated in the
following simple way: the property of a "matter" are not the same
for the different "coordinate system" is chosen.

\subsection{Equivalence frames}

The important feature of the gravity theory is connected with the
conformal symmetry. It is well known, since the pioneering paper
of Jordan \cite{Jordan} that the action is invariant under local
transformations of units that are under general conformal
transformations, or sometimes called Weyl rescaling:

\begin{equation}
ds^2 \rightarrow d\tilde{s}^2 = e^{2\alpha(x)}ds^2. \label{BDeq2}
\end{equation}
where $\alpha(x)$ a local arbitrary function of $x$.

 This method of conformal transformation provides a clear and powerful technique, free
from mathematical ambiguity, but nevertheless requires careful
consideration from the physical point of view.

Among all conformally related frames one distinguishes two frames:
Jordan's and Einstein's. Note that, unless a clear statement of
what is understood by "equivalence of frames"- is made, the issue
which is the physical conformal frame is a semantic one. For
example, by shifting a mass terms in Lagrangian one can construct
four related but inequivalent
 theories in Jordan and Einstein frame \cite{Quiros}.

 In the literature, the physicists do not agree with
each other about the equivalence of the two frames (see review in
\cite{FM}). However, the meaning of the equivalence between the
Jordan frame and the Einstein frame is not assuming the additional
equations \Ref{FeqI}. These equations put by hand and not
covariant.  This issue is critical for the interpretation of the
predictions of a given theory of gravity since these seem to be
deeply affected by the choice of the coordinate conditions
\cite{Gullstrand}. For concreteness, let us consider "harmonic
gauge" \cite{Fock},

\begin{eqnarray}\label{CHarm}
g_{\mu\nu} \Gamma^{\lambda}_{,\mu\nu}=0.
\end{eqnarray}
which usually assumed as the analogue of Lorenz gauge, $\partial
A=0$, in electromagnetism. However this analogy is the most
superficial: this or that gauge in nonrelativistic theory is a
problem of exclusively convenience, it's this or that expedient
does not influence in any way on a values of physical quantities
and it is not related to observation requirements, - whereas the
choice of co-ordinate system is related to all it essentially.

In fact, there are the related but inequivalent theories in Jordan
and Einstein frame. The reason is very simple. If we use the same
conformal transformations, like the \Ref{Conf}, in both the
equations \Ref{EquE} and \Ref{FeqI}, then the in and out states
are not the same in the two frames.  If one postulates that the
field equations are invariant with respect to conformal
transformations \Ref{FeqI}, one obtains in addition
transformations of co-ordinate conditions 

\begin{eqnarray}\label{CHarm}
g_{\mu\nu} \Gamma^{\lambda}_{,\mu\nu}=\tilde{g}_{\mu\nu}
\tilde{\Gamma}^{\lambda}_{,\mu\nu}+\partial_\mu \alpha.
\end{eqnarray}
As a result, since the Einstein field equations are undetermined;
gravity theory cannot achieve the harmonic metric for any $\alpha$
functions but only when $\alpha$  is taken a constant. One must
assume that two frames represent not the same set of physical
gravitational and non-gravitational fields. In fact, two
conformally connecting spaces $V^4(g)$ and $\tilde{V}^4(g)$ are
given not in the same manifold. Consequently, under this conformal
transformation the solution of some initial physical problem will
be transformed onto a solution of a completely different problem.
Thus, applying the same coordinate conditions in different
physical requirements, we arrive at dissimilar physical theories,
because we are solving different equations.

On the other hand each scalar-tensor theory can be considered as
general relativity plus conformally invariant scalar fields
\cite{Sokolowski}. The gravitational interaction for scalar tensor
theories is taken into account by the Einstein equations, which
are generally written in the form  \Ref{EquE}. The left-hand-side
of equation is constructed from the geometrical properties of the
space-time, while $T_{\mu\nu}$  is the energy momentum tensor of
matter fields. One can in principle assume gauge-dependence of
right-hand-side of equation \Ref{EquE} as a variety of matter
fields with different equations of state. Now, if we consider, the
system \Ref{EquE}, \Ref{FeqI} as equations for same "gauge" fixing
then the Jordan's and Einstein's conformal frames can be viewed as
a different "matter source" of energy momentum tensors
$T_{\mu\nu}$ of Einstein's equations.

It is evident that in different conformal frame representations
are neither mathematically, nor physically equivalent.

\section{Matching of spacetimes}

There are the relationship between the "gauge freedom" of General
Relativity and the hole argument. Actually standard Einstein hole
argument can be written by reverting to spacetime model of perfect
fluid configuration with vacuum background.
An important aspect in General Relativity is the
analysis of how to match two spacetimes. Obviously, there are
infinite ways to identify the manifolds, all of them equally valid
a priori. This freedom leads to the gauge dependence of the
emerged spacetime and of any other geometrically defined tensors.
In particular, the matched spacetime cannot be thought to exist
beforehand. Another aspect is that the matching conditions involve
exclusively tensors on the identified boundary -- and hence any
coordinate system in both spacetimes is equally valid.  Most of
the difficulties arise from the fact that the matching conditions
are imposed in specific coordinate systems.
The matching involves finding an
identification of the boundary and that this should not be fixed a
priori and fields to be matched are gauge dependent too. 

From the mathematical point of view, the most simple and
satisfactory expression for the matching conditions is, following
Linchnerowicz, the assumption that there exists a system of
co-ordinates in which the metric tensor satisfies the continuity
conditions. Let $\xi^i $ be a coordinate system on $\Sigma$ where
$\Sigma$ is an abstract copy of any of the boundaries. Greek
indices range over the coordinates of the 4-manifold and Roman
indices over the coordinates of the 3-surfaces. Continuity
conditions  require a common coordinate system on $\Sigma$ and
this is easily done if one can set $\xi_+^i
= \xi_-^i$. 

Let $(V^{\pm},g^{\pm})$ be four-dimensional spacetimes with
non-null  $\Sigma^{\pm}$. The junction/shell formalism constructs
a new manifold $\cal M$ by joining one of the distinct parts of
$V^+$ to one of the distinct parts of $V^-$ by
the identification $\Sigma^+ = \Sigma^- \equiv \Sigma$. The matching conditions 
require the equality of the first and second
fundamental forms on $\Sigma^{\pm}$.
Tangent vectors to $\Sigma^{\pm}$ are obtained by $ e^{\pm
\alpha}_i = \frac{\partial x_\pm^\alpha}{\partial \xi^i} $. There
are also unique (up to orientation) unit normal vectors $
n_{\pm}{}^{\alpha}$ to the boundaries. We choose them so that if
$n_{+}{}^{\alpha}$ points towards $V^{+}$ then $n_{-}{}^{\alpha}$
points outside of $V^{-}$ or viceversa.  Clearly the sign of the
normal vectors are crucial since e.g. $n^-_\alpha$ points away
from the portion of $V^-$ which will be used in forming $\cal M$.
The three basis vectors tangent to $\Sigma$ are
\begin{equation}
e^\alpha_{i} = \frac{\partial x^\alpha}{\partial \xi^i}
\end{equation}
which give the induced metric  (first fundamental form) on
$\Sigma$ by
\begin{equation}
q_{i j} = \frac{\partial x^\alpha}{\partial \xi^i}
          \frac{\partial x^\beta} {\partial \xi^j} g_{\alpha \beta}.
\end{equation}
The extrinsic curvature (second fundamental form) is given by
\begin{eqnarray}
\label{eqn-Kij1} K_{i j} & = & \frac{\partial x^\alpha}{\partial
\xi^i}
              \frac{\partial x^\beta}{\partial \xi^j} \nabla_\alpha
n_\beta      \nonumber    \\
\label{eqn-Kij2} & = & -n_\gamma \left( \frac{\partial^2
x^\gamma}{\partial \xi^i \partial \xi^j} + {\Gamma^\gamma}_{\alpha
\beta}
  \frac{\partial x^\alpha}{\partial \xi^i}
  \frac{\partial x^\beta}{\partial \xi^j}
  \right).
\end{eqnarray}


Then matching conditions are simply
\begin{equation}
  q_{ij}{}^{+}=q_{ij}{}^{-},
\label{eq:backm}
\end{equation}
\begin{equation}
  K_{ij}{}^{+}=K_{ij}{}^{-}.
\label{eq:backmK}
\end{equation}

If both \Ref{eq:backm} and \Ref{eq:backmK} are satisfied we refer
to $\Sigma$ as a boundary surface. If only \Ref{eq:backm} is
satisfied then we refer to $\Sigma$ as a thin-shell.

Consider be a bounded, closed spacetime region $V^{-}$ on which
the metric field $g^-$ is the only one present, so that inside
$V^{-}$, the metric $g^-$ obeys the Einstein's field equations
\Ref{EquE}. Given a solution $g^-(x)$ everywhere inside of and on
the boundary of $V^{-}$, including all the normal derivatives of
the metric up to any finite order on that boundary, this data
still does not determine a unique solution outside $V^{-}$,
because an unlimited number of other solutions can be generated
from it by those diffeomorphisms that are identity inside $V^{-}$,
but differ from the identity outside $V^{-}$. The resulting metric
$g^+(x)$ will agree with $g^-(x)$ inside of and on the boundary of
$V^{-}$, but will differ from it outside $V^{-}$.




\subsection{Matching of incompressible liquid
sphere with vacuum background}

As is well known, static solutions of Einstein's equations with
spherical symmetry (the exterior and interior Schwarzschild
solutions) are staples of courses in general relativity. In the
following analysis we assume a perfect fluid incompressible liquid
sphere as a simplest model for matter field.

 Writing two static spherically symmetric spacetimes
$V^+$ and $V^-$ with signature $(- + + + )$. one can   suppose
that the metrics $g^+_{\alpha \beta}(x_+^\gamma)$ and $g^-_{\alpha
\beta}(x_-^\gamma)$ in the coordinate systems $x_+^\gamma$ and
$x_-^\gamma$ are of the forms
\begin{equation}
ds^2=-B^{-}(r,t)dt^2 + A^{-}d(r,t) dr^2 + R^{-}(r,t) \left(
d\theta^2+\sin ^2\theta d\varphi ^2 \right) \label{eq:metricM}
\end{equation}
and
\begin{equation}
ds^2=-B^{+}(r,t)dt^2 + A^{+}d(r,t) dr^2 + R^{+}(r,t) \left(
d\theta^2+\sin ^2\theta d\varphi ^2 \right) \label{eq:metricP}
\end{equation}
where $ A^{\pm}(r,t), B^{\pm}(r,t)$ and $R^{\pm}(r,t)$ are of
class $C^2$.


Within these spacetimes define two non-null 3-surfaces $\Sigma^+$
and $\Sigma^-$ with metrics $q^+_{ij}(\xi_+^k)$ and
$q^-_{ij}(\xi_-^k)$ in the coordinates $\xi_+^k$ and $\xi_-^k$
which decompose each of the 4-spacetimes into two distinct parts.
The parametric
equation for $\Sigma$ is of the form
\begin{equation}
\label{eqn-surf} r - r_b =0. 
\end{equation}

The induced metric on the  $\Sigma$ by the two solutions
\Ref{eq:metricM} and \Ref{eq:metricP} is
\begin{equation}
q^{\pm}_{i j} d \xi_{\pm}^i d \xi_{\pm}^j = -B^{\pm}(r,t)dt^2 +
R^{\pm}(r,t) \left( d\theta^2+\sin ^2\theta d\varphi ^2 \right)
\label{eq:metricIM}
\end{equation}
therefore we must have on the $\Sigma$ according to equality of
the first fundamental form
\begin{equation}
B^{-}(r_b,t)= B^{+}(r_b,t), C^{-}(r_b,t)= R^{+}(r_b,t)
\label{eq:tirstFF}
\end{equation}

 We can impose that the $\Sigma$ be also
characterized by equality of the second fundamental form too. This
condition leads to
\begin{equation}
K_{1 1} = \frac{1}{2}\frac{B'(r_b,t}{\sqrt{A(r_b,t)}}, K_{2 2} =
 \frac{1}{2}\frac{R'(r_b,t)}{\sqrt{A(r_b,t)}} \label{eq:tirstSF}
\end{equation}
where prime denote derivation wits respect to $r$. Continuity of
the second fundamental form  is merely equivalent to

\begin{eqnarray}
 \!\!\!\!\! B^{-}(r_b,t)= B^{+}(r_b,t), R^{-}(r_b,t)= R^{+}(r_b,t), \nonumber \\
R^{+'}(r_b,t)^2 A^{-} = R^{-'}(r_b,t)^2 A^{+},B^{+'}(r_b,t)^2
A^{-} = B^{-'}(r_b,t)^2 A^{+}
 \label{eq:MC}
\end{eqnarray}

 Note that in the case
of incompressible liquid model with curvature coordinates the
requirement of matching spacetimes embraces continuity of the
metric function. However, the derivative of $g_{1 1}$ is
inescapably discontinues, and the derivative of $g_{0 0}$ is
already continuous without the necessity of requiring it
\cite{Aguirregabiria}. This is of course the same that happens
when trying to match other exterior and interior spherically
symmetric solutions. The solution of Einstein equation for
\Ref{eq:metricM} can be reduced to

\begin{equation}
d s^2 = - \frac{(1+r^2/S^2)^2}{1+ r^2/r_b^2)^2}  d t^2 + \frac{d
r^2 + r^2 (d \theta^2 + \sin^2 \theta d \phi^2)} {(1+
r^2/r_b^2)^2}.\label{eq:const}
\end{equation}
A brief computation yields
\begin{equation}
\rho= {12\over r_b^2}; \qquad p=  {4\over S^2 r_b^4}
{r_b^2(r_b^2-2S^2)-(2r_b^2-S^2)r^2 \over1+r^2/S^2 }.
\end{equation}
where $S$ arbitrary constant.

After finding interior solutions, we can then connect them to the
exterior vacuum solutions. We take $r_b$ to be the point where
$p(r) = 0$, and use the values of $A^-(r_b), B^-(r_b), C^-(r_b),
A^{-'}(r_b), B^{-'}(r_b), C^{-'}(r_b)$ from the solutions as
conditions to determine the unknown integration coefficients from
the vacuum case. As we pointed out unlimited number of spherically
symmetric vacuum solutions can be obtained outside $V^{-}$. It is
easy to show, however, that we can match the interior solution
with the most general spherically symmetric vacuum solution
\cite{Logunov}
\begin{equation}
d s^2 = \frac{\rho^{+'}(r)^2}{4 \rho^{+}(r)} \left(1-\frac{2
\mu}{\sqrt{\rho^{+}(r)}}\right)^{-1}  d t^2 + (1-\frac{2
\mu}{\sqrt{\rho^{+}(r)}})d r^2 +\rho^{+}(r) \left(d \theta^2 +
\sin^2 \theta d \phi^2\right).\label{eq:vacuum}
\end{equation}
where $\rho^{+}(r)$ is a arbitrary function of $r$.  
To perform the matching, we impose the arbitrary function
$\rho^{+}(r)$ in a polynomial form. The conditions \Ref{eq:backm}
must be fulfilled at the boundary where $p = 0$. These continuity
conditions provide us with the information needed to find values
for the arbitrary polynomial coefficients for the vacuum
solutions. After matching the solutions in this manner, $B'(r)$
is not necessarily continuous at the boundary. 



Now in order to justify calling the geometry an exact solution we
need an explicit definition for the constant in these solutions.
The integration constants of solution \Ref{eq:const} and
\Ref{eq:vacuum} are arbitrary. Obviously, it is possible to match
the solution \Ref{eq:const}  to the vacuum \Ref{eq:vacuum} metric
with unlimited number of arbitrary function $\rho^{+}(r)$.

 Finally, we observed some surprising ambiguity of interpretation to
 a choice of function $\rho^+(r)$ generally corresponds to a choice of
state within the vacuum. For example, the solution \Ref{eq:vacuum}
with  different choose of arbitrary function to $\rho^{+}(r)$ have
the same spatial boundary behavior but have different property,
and so they represent different vacuum states. Thus, from the
point of view that the field equations \Ref{EquE} is just a formal
device to arrive at the space time, information about the possible
vacua of the theory, and the space of states in each vacuum, is
not encoded directly in the \Ref{EquE}, only indirectly through
the "gauge" fixing and boundary conditions required of the
equations.



It is seen by inspection that all junction conditions \Ref{eq:MC}
are satisfied. We can therefore say that spacetime metric
generated by static spherically symmetric distribution of perfect
fluid incompressible matter can be matched with wide family of
vacuum solution by suitable choose the integration constants or
vice versa. We therefore can to see what any explicit constraints
on the exterior spacetime geometry could not be deduced directly
from the interior perfect fluid solution, without reference to a
"gauge" fixing or viceversa.
\section{Conclusion}

In this article, we clarify the notion of arithmetization,
"gauges" and coordinate transformations in relativistic theories,
which is necessary to understanding the physical equivalence
between the different coordinate system. Einstein Hilbert action
is ambiguously decomposed into the sum of physical term, which
represent the gravitational effects, and a pure geometrical term
which represent the spurious gravitational effects associated with
manyfold arithmetization. The matter terms of the Lagrangian
density contain the connection and hence a part of dynamical
description of gravity, that is invariant in form under the
conformal "rescaling" of the metric. In other words the matter
terms  may be represented in disguise in infinite number of way as
a gravity component or as a conformally invariant fields. Once
matter has been coupled to gravity in a frame one has a freedom to
make a change of variables to any other frame. So property of
Einstein Hilbert action in general relativity is indeed an
observer dependent concept.

In relativistic theory, we must always write in addition to fields
equation \Ref{EquE} four co-ordinate conditions \Ref{FeqI}. These
"gauges" may describe different physical solutions of Einstein
equations with the same space arithmetization.  We have shown the
"gauge" fixing influences results of measurements and physics are
different in two different coordinate system.

The gravitational interaction are described by a doublet
constituting of a metric tensor and a matter fields, the important
feature of general relativity is connected with conformal
symmetry. In particular, according to this view, general
relativity may be rewritten in terms an
arbitrary conventional geometry \cite%
{Tavakol} and the geometry of space-time can be freely chosen by
the theoretician \cite{Henri}. In method of conformal
transformation, we always treat two spacetimes. First is the
space-time for one frame and the other is the space-time for
another frame. Note that the two space-times for these frames are
distinct. The conformal transformations are not diffeomorphisms of
the single manifold $M$, and the transformed metric
$\tilde{g}_{\mu\nu}$ is not simply the metric $g_{\mu\nu}$ written
in a different coordinate system these metrics describe different
gravitational fields and different
physics. 

Eq. \Ref{BDeq2} is a rather curious equation because it not
covariant for the arbitrary transformations of independent
variables. In this case the metric is left unchanged, although its
coordinate representation varies. In short, Eq. \Ref{BDeq2} gives
a relation between variables on two different
space-times.

Evidently, a structure of space-times can be mathematically
represented with cosmological and coupling \emph{constants}; the
conformally changed Einstein equations have the advantage of
non-vanishing modified terms together with \emph{dynamical}
cosmological and gravitational coupling terms.

The Einstein's hole argument states a spacetime and a
gravitational field form an indivisible unit: no field, no
spacetime. We consider the vacuum static spherically symmetric
solutions of general relativity to illustrate this. One might
represent the metric tensor components and it's first derivatives
on a boundary hypersurface \Ref{eq:const} with some constrains
from the field equations, would uniquely determine the solution in
neighborhood spacetime. But no such boundary condition can do
this: any solution \Ref{eq:vacuum} can be transformed to other by
a suitable choose of arbitrary function $\rho(r)$. The field
equations cannot even uniquely determine the geometry of a
spacetime on which a solution is defined.

\end{document}